\documentclass[prb,aps,onecolumn,floatfix]{revtex4}
\usepackage[pdftex]{graphicx}
\usepackage{bm}
\usepackage{xcolor}
\usepackage{wasysym}
\usepackage{mathrsfs}

\renewcommand{\deg}{$^{\circ}$}
\newcommand{\etal}{ {\it et al.}}
\newcommand{\newc}{\newcommand}
 
\newc{\be}{\begin{equation}}
\newc{\ee}{\end{equation}}
\newc{\bfe}{\begin{floatequation}}
\newc{\efe}{\end{floatequation}}
\newc{\bea}{\begin{eqnarray}}
\newc{\eea}{\end{eqnarray}}
\newc{\ie}{{\it i.e.} }
\newc{\eg}{{\it e.g.} }
\newc{\etc}{{\it etc.} }
\newc{\ra}{\rightarrow}
\newc{\lra}{\leftrightarrow}
\newc{\lsim}{\buildrel\langle\over{\sim}}
\newc{\gsim}{\buildrel\rangle\over{\sim}}

\newc{\one}{\mathbbm{1}}
\newc{\Tr}[1]{\mathrm{Tr}\left[ {#1} \right]}
\newc{\ket}[1]{\left|{#1}\right\rangle}
\newc{\bra}[1]{\left\langle{#1}\right|}
\newc{\braket}[2]{\langle{#1}|{#2}\rangle}
\newc{\mean}[1]{\langle{#1}\rangle}
\newc{\braketd}[1]{\langle{#1}|{#1}\rangle}
\newc{\ketbrad}[1]{\left|{#1}\rangle\!\langle{#1}\right|}
\newc{\ketbra}[2]{\left|{#1}\rangle\!\langle{#2}\right|}
\newc{\EV}[2]{\langle{#1}\rangle_{#2}}
\newc{\C}{\ensuremath{\mathbbm C}}

\def \bmlett{\begin{mathletters}}
\def \emlett{\end{mathletters}}
\def \r{{\bf r}}

\def \ra{\rightarrow}

\def\be{\begin{equation}}
\def\ee{\end{equation}}

\def\w01{\omega_{01}}
\def\r0{R_0}

\begin{document}

\title{Angular Preisach analysis of Hysteresis loops and FMR lineshapes of ferromagnetic nanowire arrays} 

\author{C. Tannous, A. Ghaddar}

\affiliation{Laboratoire de Magn\'etisme de Bretagne - CNRS FRE 3117- Universit\'e de Bretagne Occidentale -
6, Avenue le Gorgeu C.S.93837 - 29238 Brest Cedex 3 - FRANCE.}

\author{J. Gieraltowski}
\affiliation{Laboratoire des domaines Oc\'eaniques, IUEM CNRS-UMR 6538, Technopole Brest IROISE 29280 Plouzan\'e, FRANCE.}

\begin{abstract}
Preisach analysis is applied to the study of hysteresis loops measured for different 
angles between the applied magnetic field and the common axis of ferromagnetic Nickel 
nanowire arrays. When extended to Ferromagnetic Resonance (FMR) lineshapes, 
with same set of parameters extracted from the corresponding hysteresis loops,
Preisach analysis shows that a different distribution of interactions or coercivities
ought to be used in order to explain experimental results. Inspecting the behavior of
hysteresis loops and FMR linewidth versus field angle, we infer that angular dependence 
might be exploited in angle sensing devices that could compete with 
Anisotropic (AMR) or Giant Magnetoresistive (GMR) based devices.
\end{abstract}

\maketitle

\section{Introduction}

Ferromagnetic nanowires possess interesting properties that might be exploited
in spintronic devices such as race-track type magnetic non-volatile memory called MRAM  
(based on transverse domain-wall dynamics~\cite{Dynamics,Torque}) 
and magnetic logic devices~\cite{Sun,Yan,Katine}. They might also be used in 
magnonic devices based on spin-wave excitation and propagation~\cite{magnon}. 

Ferromagnetic nanowires have applications 
in microwave devices such as circulators~\cite{coupling}, superconducting single-photon
GHz detectors and counters~\cite{counters}, information storage (as recording media
and read-write devices), Quantum transport (such as GMR~\cite{GMR} circuits)
as well as in Quantum computing and Telecommunication.

They are simpler than nanotubes since their physical properties do not
depend on chirality and they can be grown with a variety of methods~\cite{Growth,Tannous}: Molecular
Beam Epitaxy, Electrochemical methods (Template synthesis, Anodic Alumina filters), 
Chemical solution techniques (Self-assembly, Sol-Gel, emulsions...) and can be grown 
with a tunable number of monolayers and length~\cite{Tunable}.

Ordered arrays of nanowires may be of paramount importance in areas  
such as high-density patterned media information recording an example of which is the
Quantum Magnetic Disk~\cite{Chou}. They might be also of interest in novel 
high-frequency communication or signal-processing devices based on the
exploitation of spin-waves (in magnonic crystals made of magnetic superlattices
or multilayers)~\cite{magnon} to transfer 
and process information or spin-currents with no dissipative Joule effect.

In this work, we explore the possibility for Nickel ferromagnetic nanowire arrays 
(FNA) to be of interest in angle sensitive devices. For this goal we perform field angle dependent 
hysteresis loops and FMR lineshape measurements  in the X-band (9.4 GHz).
Preisach analysis is applied to extract from the measured effective anisotropy field $H_{eff}$ 
several angle dependent physical parameters (such as interaction and coercivity) 
while changing nanowire diameter from 15 nm to 100 nm. 

These findings might be exploited in angle dependent sensing devices that might 
compete with present AMR or GMR angle sensors.
 
This work is organized as follows: In section 2, measured hysteresis loops versus field angle
are presented and analyzed with Preisach modeling, whereas in section 3 the same
analysis is performed on the FMR lineshape measurements. 
We conclude the work in section 4.
Appendix I details the FMR angular fitting procedure whereas Appendix II is a 
general overview of Preisach modeling.

\section{Hysteresis loops versus field angle}
\label{Preisach}

Our Nickel FNA are fabricated with an electrochemical deposition method 
similar to the one used by Kartopu \etal~\cite{Growth}
and the common length is 6 $\mu$m for all diameters while the average interwire distance 
is about 350 nm.

We have performed angle (0\deg, 30\deg, 45\deg, 60\deg and 90\deg) 
dependent VSM (Vibrating Sample Magnetometry) 
and FMR on these variable diameter (15 nm, 50 nm, 80 nm and 100 nm) arrays from
liquid Helium (4.2 K) to room temperature~\cite{Tannous}.
We have shown that the easy axis orientation for the 
15 nm diameter sample is perpendicular to the wire axis in sharp contrast with  
the 50 nm, 80 nm and 100 nm samples. This is a surprising result
since we expect (from bulk Ni) that the easy axis along the wire axis
by comparing the value of shape energy with respect to anisotropy energy. 

\begin{figure}[!h]
\centerline{\includegraphics[width=2.5in]{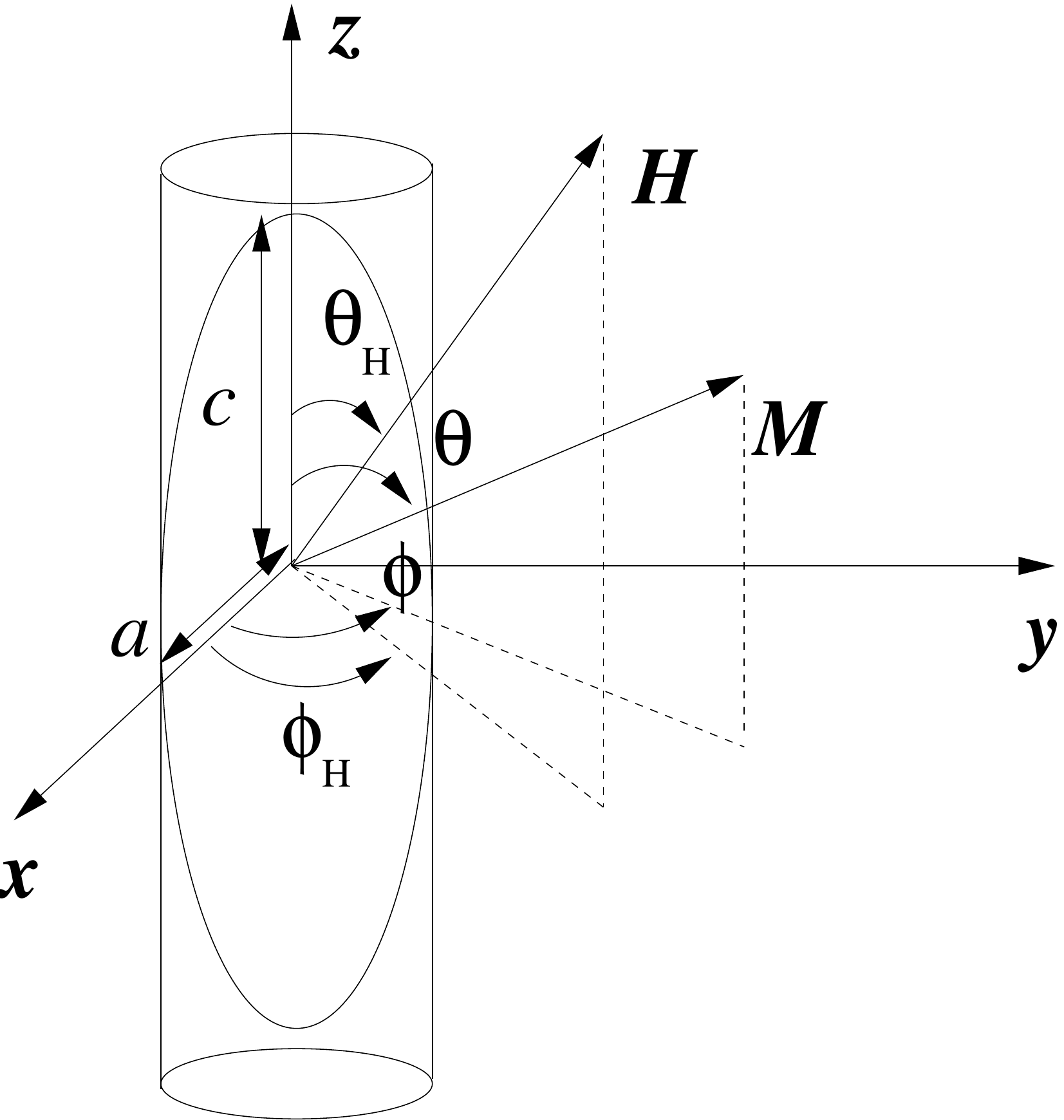}}
\caption{Magnetization $\bm{M}$, applied field $\bm{H}$ and corresponding angles 
$\theta ,\,\;\phi ,\;\,\theta_H ,\phi_H$ they make with the nanowire axis 
that can be considered as an ellipsoid-shaped single domain with characteristic lengths 
$a=d/2$ and $c$ with $d$ the diameter. When the aspect ratio $c/a$ is large enough 
the ellipsoid becomes an infinite cylinder.}
\label{coord}
\end{figure}

Results obtained from the angular behavior of the resonance field
$H_{res}$ versus $\theta _H $ shows that $H_{res}$ is minimum at 90\deg for the 15 
nm sample whereas it is minimum at 0\deg  for the larger diameter samples
agree with hysteresis loops obtained from VSM measurements and confirm presence
of the transition of easy axis direction from perpendicular at 15 nm to
parallel to nanowire axis at 50 nm diameter.

In this work we concentrate on room-temperature angle dependence 
experimental results and modeling. Preisach modeling is used to understand 
the angular behavior of the hysteresis loops and the FMR lineshapes.
After explaining the shortcomings of the Classical Preisach Model (CPM) we use
the Preisach Model for Patterned Media (PM2) to interpret the static (hysteresis loops)
and the dynamic (FMR) measurements for all angles and diameters. 

The Preisach modeling we use is based essentially on probability densities 
for interaction $h_i$ and coercive $h_c$ fields. 

If we rotate the field distribution $(h_i,h_c)$ by 45\deg with respect to
the reference system $(H_\alpha, H_\beta)$ (or switching field system; see Appendix I),
we get the relations:

\begin{equation}
h_i=(H_\alpha+H_\beta)/\sqrt{2}, h_c=(H_\alpha-H_\beta-2 H_0)/\sqrt{2} 
\end{equation}

where $H_0$ is the distribution maximum.

The CPM density is given by a product of two Gaussian densities pertaining 
to the interaction and coercive fields degrees of freedom:

\begin{equation}
p(h_i,h_c)=\frac{1}{2 \pi \sigma_i \sigma_c} \exp(-\frac{h_i^2}{2 \sigma_i^2})
\exp(-\frac{h_c^2}{2 \sigma_c^2}) 
\label{CPM}
\end{equation}

where the standard deviation of the interaction and coercive fields are given
by $\sigma_i$  and $\sigma_c$ respectively.

The PM2 model is based on the following description:

\begin{eqnarray}
 p(h_i,h_c) =\frac{1}{2 \pi \sigma_i \sigma_c} \exp(-\frac{h_c^2}{2 \sigma_c^2}) \times \hspace{2cm} \nonumber \\
\{ (\frac{1+m}{2}) \exp(-\frac{{(h_i-h_{i0})}^2}{2 \sigma_i^2}) + 
 (\frac{1-m}{2}) \exp(-\frac{{(h_i+h_{i0})}^2}{2 \sigma_i^2}) \}
\label{PM2}
\end{eqnarray}

where the normalized magnetization $m=\frac{M}{M_S}$ has been introduced as well as an average interaction
field $h_{i0}$. The magnetization $M$ is determined by double integration over the field distribution (see Appendix I).
Moreover, coercivity is represented by a single Gaussian density whereas
interactions are represented by a superposition of two Gaussian densities shifted to 
left and right with respect with respect to the average field $h_{i0}$ .

\begin{figure}
  \begin{center}
    \begin{tabular}{c}
      \resizebox{60mm}{!}{\includegraphics{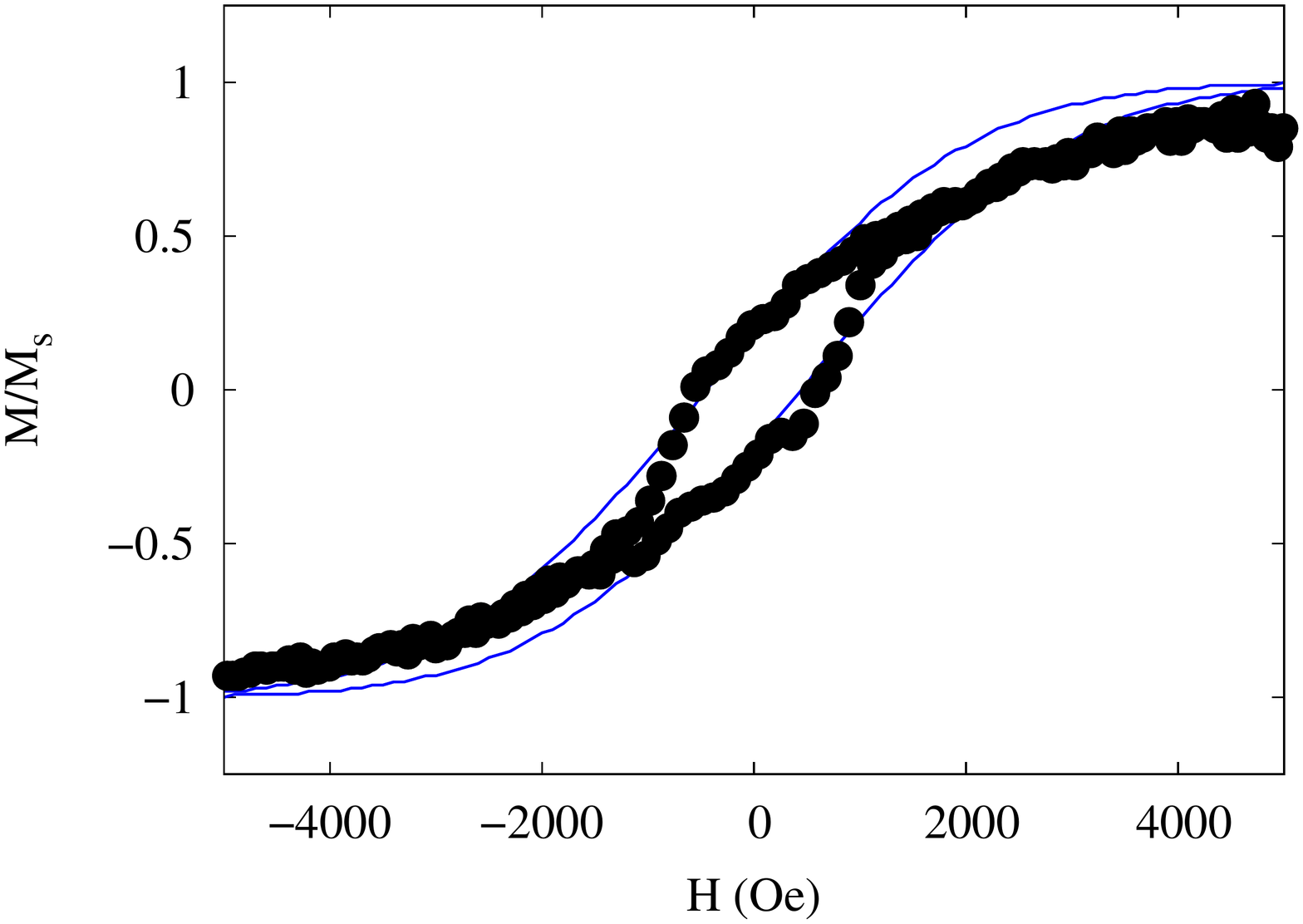}} \\
      \resizebox{60mm}{!}{\includegraphics{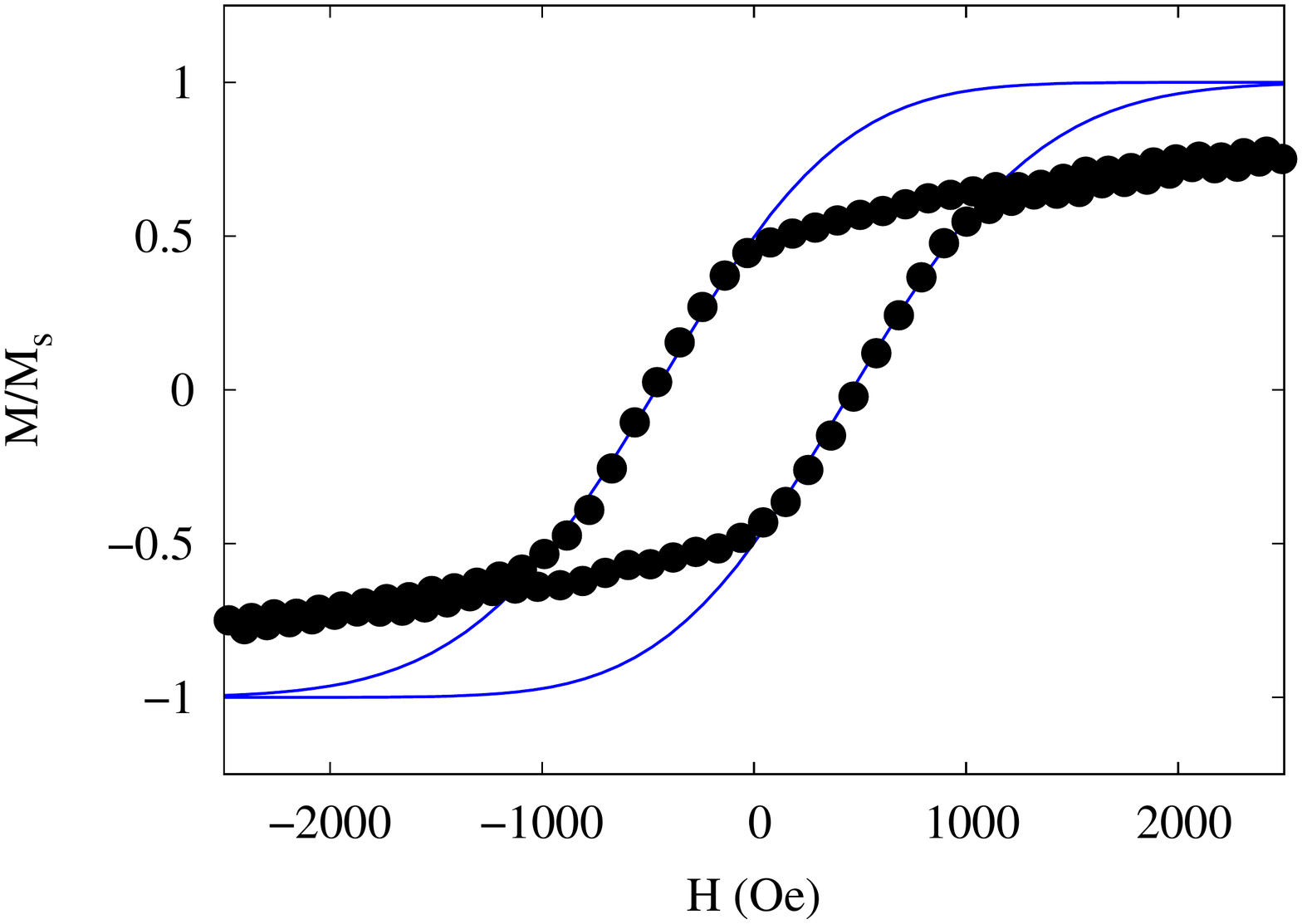}} \\ 
      \resizebox{60mm}{!}{\includegraphics{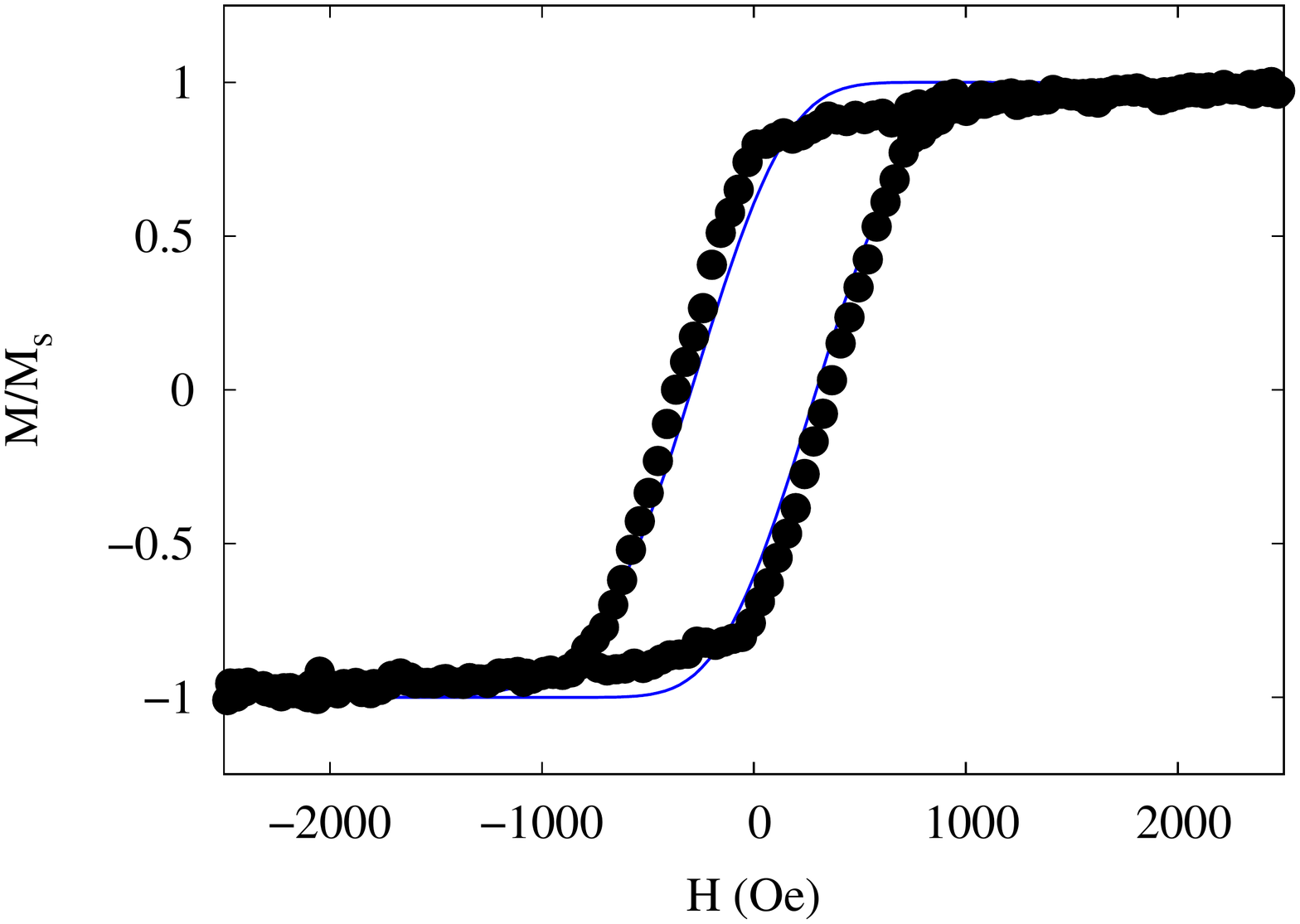}} \\
      \resizebox{60mm}{!}{\includegraphics{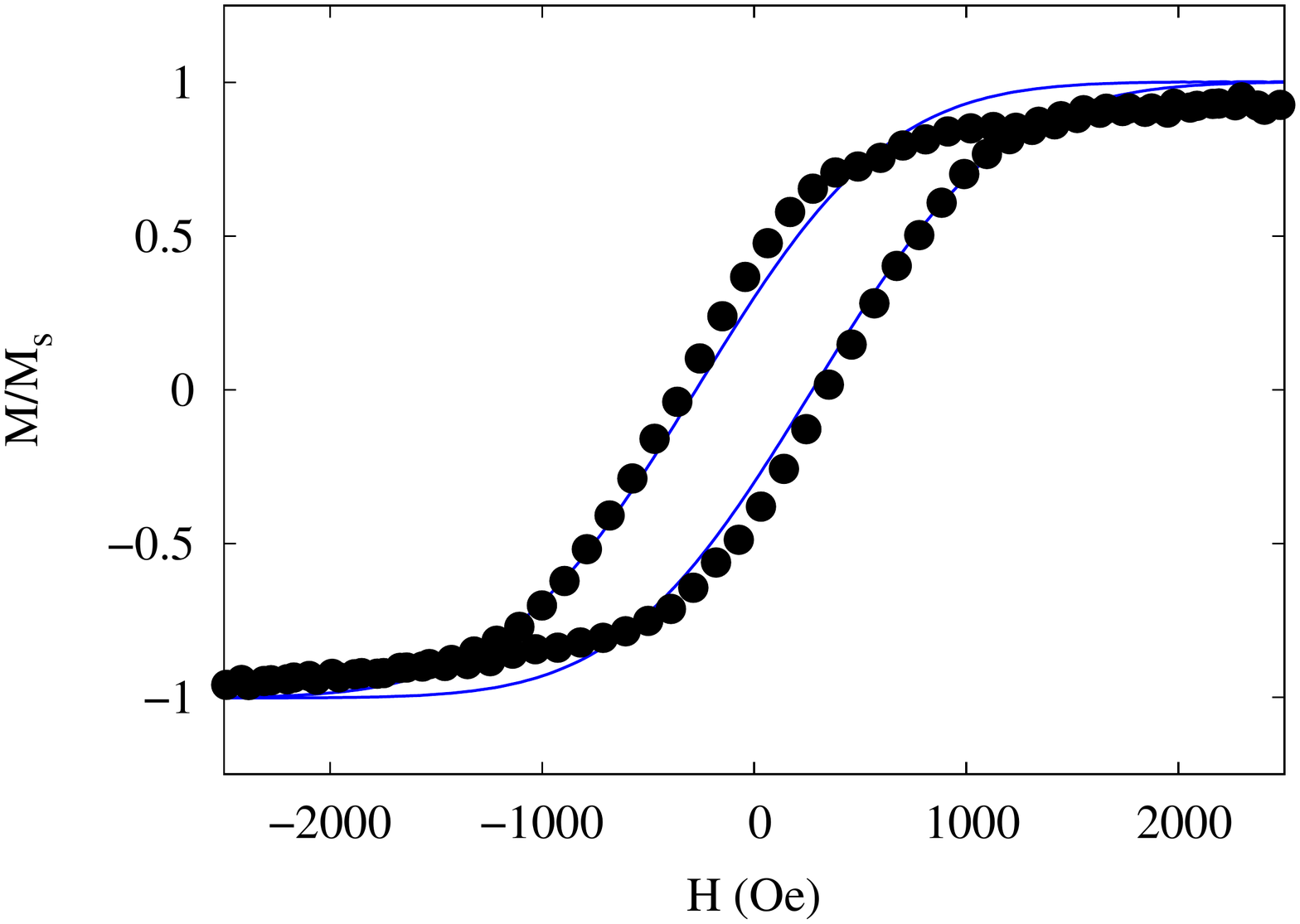}} \\ 
    \end{tabular}
    \caption{(Color on-line) Room temperature VSM measured hysteresis loops  (continuous
blue lines) $M/M_S$ versus $H$ (in Oersteds) in the 15, 50, 80 and 100 nm 
diameter cases (displayed from top to bottom) and their Preisach fit (black dots). 
The field is along the nanowire axis. Note the mismatch observed in the 50 nm case as discussed in the text.}
    \label{VSM}
  \end{center}
\end{figure}

Loop inclination increases with $\sigma_i$ whereas loop width
increases with $\sigma_c$. Hence interactions between "hysterons" (or nanowires in our case)
are responsible for the inclination observed in the VSM hysteresis loops.
The fit parameters are given in table~\ref{VSM_PM2}.

\begin{table}[htbp]
\begin{center}
\caption{Fitting parameters of the PM2 model of the hysteresis loops of Ni 
15, 50, 80 and  100 nm diameter samples when  external field $H$ is along 
nanowire axis i.e. $\theta_H=0$ (see fig.~\ref{coord}).}
\begin{tabular}{|c|c|c|c|c|}
\hline
$d$(nm) & $H_{0}$ (Oe) & $h_{i0}$ (Oe) & $\sigma_i $  (Oe) & $\sigma_c $   (Oe) \\
\hline
15 & 120 & 600 & 1800 & 555 \\ 
50 & 180 & 30 & 600 & 600  \\
80 & 170 & 170 & 200 & 390  \\ 
100 & 170 & 300 & 650 & 348  \\ 
\hline
\end{tabular} 
\label{VSM_PM2}
\end{center}
\end{table}

The coercivity parameter reflects presence of pinning centers hampering domain motions.

The Preisach fit is next made on the angular (between field and nanowire axis, see fig.~\ref{coord})
dependent hysteresis loops and we
present in Table~\ref{angle} the detailed results, as an example, for the 100 nm diameter case.

\begin{table}[htbp]
\begin{center}
\caption{Fitting parameters of the PM2 model of the hysteresis loops of Ni 
100 nm diameter samples for several angles $\theta_H$ between external field $H$ and 
nanowire axis (see fig.~\ref{coord}).}
\begin{tabular}{|c|c|c|c|c|}
\hline
Angle $\theta_H$ & $H_{0}$ (Oe) & $h_{i0}$ (Oe) & $\sigma_i $  (Oe) & $\sigma_c $   (Oe) \\
\hline
0$^{\circ}$ & 170 & 300 & 650 & 348 \\ 
30$^{\circ}$ & 170 & 200 & 750 & 360  \\
45$^{\circ}$ & 170 & 100 & 900 & 345  \\ 
60$^{\circ}$ & 100 & 50 & 1180 & 365  \\ 
90$^{\circ}$ & 100 & 20 & 1300 & 450  \\ 
\hline
\end{tabular} 
\label{angle}
\end{center}
\end{table}

One of the great advantages of the PM2 model is that the loop width is controlled by the standard 
deviation of the coercive fields $\sigma_c$ whereas its inclination is tunable by the standard 
deviation of the interaction fields $\sigma_i$. Once we fit the hysteresis loop 
we use the same parameters to evaluate the FMR lineshape as explained next.

\section{FMR lineshape versus angle}
Individual wires inside the array are aligned parallel to each other 
within a deviation of a few degrees. They are characterized by a cylindrical shape with a 
typical variation in diameter of less than 5\% with a low-surface roughness and a 
typical length of 6 microns.

FMR experiments 
are performed with the microwave pumping field $h_{rf}$ operating at 9.4 GHz
with a DC bias field $H$ making a variable angle $\theta_H$  with the nanowire axis
(through sample rotation). 

Previously, several studies have considered reversal modes by 
domain nucleation and propagation (see for instance Henry \etal~\cite{Henry} for an 
extensive discussion of the statistical determination
of reversal processes and distribution functions of 
domain nucleation  and  propagation fields). Moreover, Ferr\'e \etal~\cite{Ferre} and
Hertel~\cite{Hertel} showed the existence of domains with micromagnetic
simulations). We do not consider domain nucleation and propagation in this work and
rather concentrate on transverse single domain case.

Thus, the angular dependence of $H_{res}$ in the uniform mode is obtained by
considering an ellipsoid with energy $E$ comprised of a small second-order 
effective uniaxial anisotropy~\cite{cubic} contribution ($K_1$ term in eq.~\ref{energy})
and (shape) demagnetization energy ($\pi M_S^2$ term in eq.~\ref{energy} 
with $M_S$ the saturation magnetization). 
Their sum is the total anisotropy energy $E_A$ to which we add a Zeeman term $E_Z$ 
due to the external field $H$:

\begin{eqnarray}
E = E_A + E_Z  =  (K_1 +\pi M_S^2)\sin^2 \theta \nonumber \\
   -  M_S H [ \sin \theta \sin \theta_H \cos(\phi -\phi_H) + \cos \theta \cos \theta_H]
\label{energy}
\end{eqnarray}

$\theta$ is the angle the magnetization makes with the nanowire axis (see fig.~\ref{coord}).

The resonance frequency is obtained from the Smit-Beljers~\cite{Smit} 
formula that can be derived from the Landau-Lifshitz equation of motion
with a damping term $\alpha$:

\begin{equation}
\left[ {\frac{\omega }{\gamma }} \right]^2=\frac{(1+\alpha ^2)}{M_S^2 \sin ^2\theta }
\left[ {\frac{\partial ^2E}{\partial \theta ^2}\frac{\partial ^2E}{\partial \phi ^2}-
\left( {\frac{\partial ^2E}{\partial \theta \,\partial \phi }} \right)^2} \right] 
\label{smit1}
\end{equation}

The frequency linewidth is given by:

\begin{equation}
\Delta \omega=\frac{\gamma \alpha}{M_S}
\left( {\frac{\partial ^2E}{\partial \theta ^2}+
 \frac{1}{\sin ^2\theta } \frac{\partial ^2E}{\partial \phi ^2}} \right)
\label{smit2}
\end{equation}

The frequency-field dispersion relation is obtained from the Smit-Beljers equation after 
evaluating the angular second derivatives~\cite{Ebels} of the total energy and taking $\phi=\phi_H=\frac{\pi}{2}$:
\begin{eqnarray}
\frac{\omega}{\gamma}=\sqrt{(1+\alpha ^2)[ H_{{eff}} \cos 2\theta + H \cos(\theta -\theta_H)] }  \times \nonumber \\
 \hspace{3cm}  \sqrt { [  H_{{eff}} \cos^2\theta + H \cos(\theta -\theta_H)] }
\label{smit3}
\end{eqnarray}

At resonance, we have $\omega=\omega_r$, the resonance frequency, 
$\theta =\theta_H$ and the applied field $H=H_{res}$, the resonance field,
when we are dealing with the saturated case. 
In the unsaturated case the magnetization angle $\theta \neq \theta_H$ and one
determines it directly from energy minimization.

The above relation~\ref{smit3} provides a relationship between the 
effective anisotropy field 
$\bm{H}_{eff}$ and the external field $H$ at the resonance frequency.

Generally, the effective anisotropy field $\bm{H}_{eff}$ can be obtained 
from the vectorial functional derivative of the energy $E_A$ (eq.~\ref{energy}) with respect to
magnetization $\bm{H}_{eff}~=~-\frac{\delta E_A}{\delta \bm{M}}$ that becomes
in the uniform case the gradient with respect to the magnetization components 
$\bm{H}_{eff}~=~-\frac{\partial E_A}{\partial \bm{M}}$.

Moreover, we need to determine magnetization orientation $\theta _0$ at equilibrium. 
This is obtained from the minimum condition by evaluating the first derivative 
${(\frac{\partial E}{\partial \theta })}_{\theta_0}=0$
and requiring positivity of the second derivative. 
Consequently, we get:

\begin{equation}
(K_1 +\pi M_S^2) \sin 2\theta_0 =M_S \,H\,\sin \,\left( {\theta _H -\theta _{0} } \right)
\label{min}
\end{equation}

This equilibrium equation \ref{min} and Smit-Beljers equation~\ref{smit3} are 
used simultaneously to determine the resonance field 
$H_{res}$ versus angle $\theta_H$ as analyzed next.

\subsection{Analysis of the effective anisotropy field}

In order to evaluate the total effective anisotropy field $H_{{eff}}$,
we include in the energy $E_A$, demagnetization, magnetocrystalline anisotropy,
and interactions among nanowires, with the corresponding fields $H_{dem}=2 \pi M_S$,
$H_K=\frac{2 K_1}{M_S}$ and $H_{i}$, thus:

\begin{equation}
H_{{eff}} =H_{dem} +H_{i} +H_K 
\label{heff}
\end{equation}

The interaction field $H_{i}$ comprises dipolar interactions between nanowires that 
depend on porosity $P$ (filling factor) and additional interactions as described
in the CPM and PM2 models (see Appendix II). For example, if we consider
the simplest case, demagnetization and dipolar fields are of the same form 
and may be written~\cite{Ebels} as a single term $2 \pi M_S (1-3P)$.

Experimentally,  the resonance field $H_{res}$ versus field angle $\theta_H$ peaks~\cite{Encinas} 
at $\omega_r/\gamma $, hence
it is possible to extract the effective anisotropy field $H_{{eff}}$ through the use
of eq.~\ref{smit3}. Thus the Land\'e
$g$-factor, saturation magnetization $M_S$ and cubic anisotropy constant $K_1$ can be determined
with a least-squares fitting method~\cite{Tannous} similar to the one used in Appendix I.

This yields the following table~\ref{fit_table} containing fitting parameters 
$K_1$ and $M_S$ (Anisotropy and saturation magnetization) versus diameter.

\begin{table}[htbp]
\begin{center}
\caption{Room temperature fitting parameters $K_1$ and $M_S$ with corresponding 
Nickel nanowire diameter $d$ and average separation $D$. 
Effective $H_{{eff}}$ and anisotropy $H_{K}$ fields are determined with Smit-Beljers.
Comparing with bulk Nickel anisotropy~\cite{aniso} coefficient at room temperature: 
$K_1=-4.5 \times 10^{4}$ erg/cm$^3$ and saturation magnetization  
$M_S$=485 emu/cm$^3$ we infer that as the diameter increases we get 
closer to the bulk values as expected with $K_1$ changing by about 
two orders of magnitude.}
\begin{tabular}{|c|c|c|c|c|c|}
\hline
$d$ & $D$ & $K_1$  & $M_S$  & $H_{{eff}} $   & $H_{K}$    \\
(nm) & (nm) & (erg/cm$^3$) & (emu/cm$^3$) & (Oe) & (Oe) \\
\hline
15& 256    &  -1.909 $\times$ 10$^{6}$  &  988.22 & 2344.58   &  -3864.61  \\
50& 510  & -1.621 $\times$ 10$^{5}$   & 451.95 &   2122.17    &  -717.52 \\
80& 393   &  -2.424 $\times$ 10$^{5}$ &    453.25 &  1778.32   &  -1069.56 \\
100& 497  &  -8.037 $\times$ 10$^{4}$    &   410.24  &   2185.78  &  -391.81 \\
\hline
\end{tabular}
\label{fit_table}
\end{center}
\end{table}

From table~\ref{fit_table}, one infers that as the diameter increases the Ni bulk values are steadily
approached which is a good test of the FMR fit.

\subsection{Preisach modeling of FMR lineshape and transverse susceptibility}

The FMR lineshape is obtained from the field derivative  $\frac{d <\chi_{xx}^{''}>}{dH}$
of the average  transverse susceptibility imaginary part  $<\chi^{''}_{xx}>$  given by:

\begin{equation}
<\chi_{xx}^{''}>=\iint_S p(H_\alpha, H_\beta) \chi_{xx}^{''} dH_\alpha dH_\beta
\label{chi}
\end{equation}

The fields $H_\alpha, H_\beta$ are the switching fields that define the Preisach plane (see Appendix II)
over which the double integration above is performed in order to estimate the average.

The expression of the transverse susceptibility imaginary part is derived directly from 
the energy \cite{Dumitru} and given by:

\begin{eqnarray}
\chi_{xx}^{''}=  \frac{\omega}{{(\omega_r^2 - \omega^2)}^2 + \omega^2 \Delta \omega_r^2} \times \nonumber \\
 \left[ -\gamma^2 (1+\alpha^2) \left(\frac{\partial^2 E}{\partial \theta^2}\right) \Delta \omega_r + \alpha \gamma M_S (\omega_r^2 - \omega^2)  \right]
\label{chixx}
\end{eqnarray}

\begin{figure}[htbp]
\begin{center}
    \begin{tabular}{c}
      \resizebox{80mm}{!}{\includegraphics[angle=0]{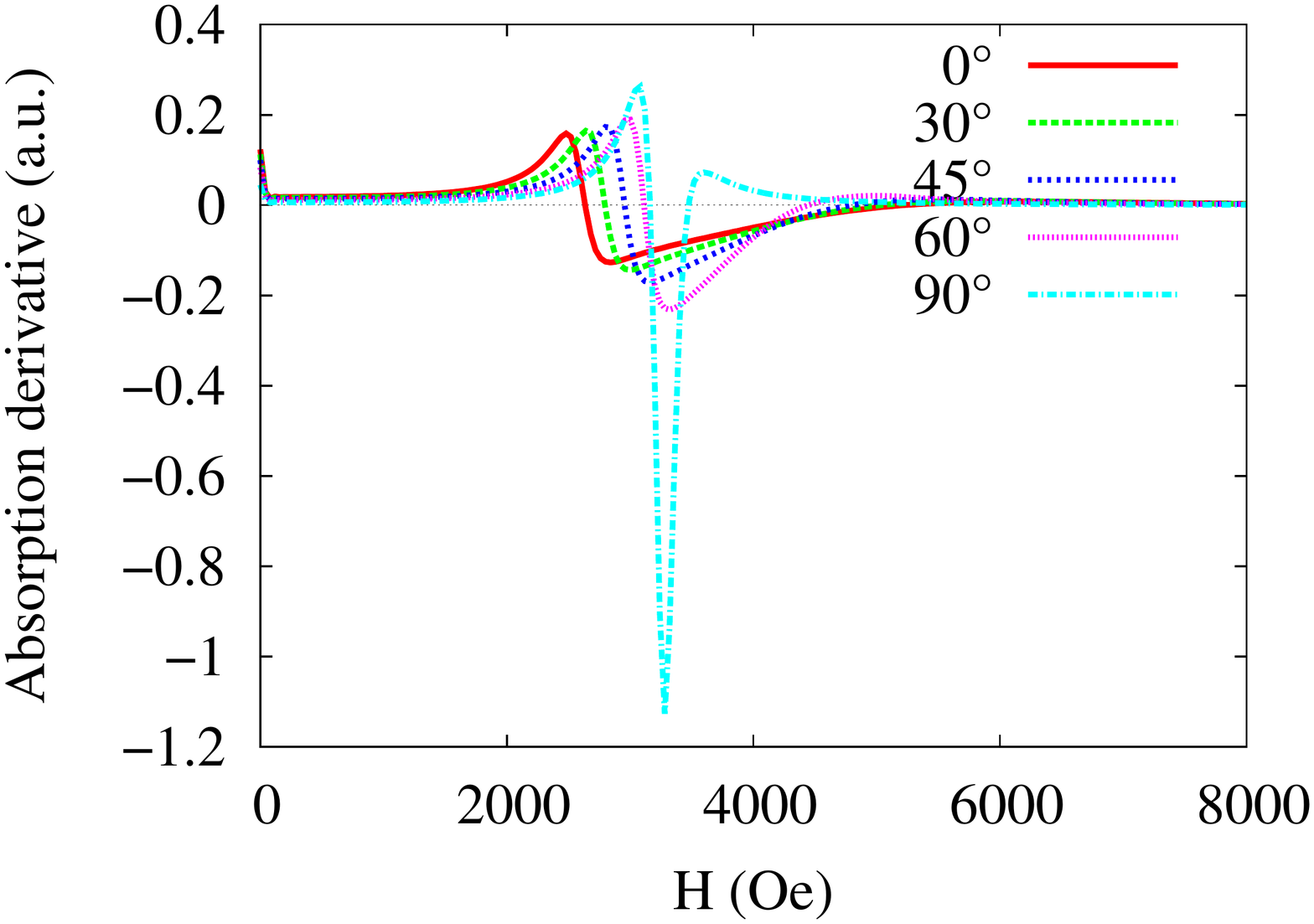}} \\
      \resizebox{80mm}{!}{\includegraphics[angle=0]{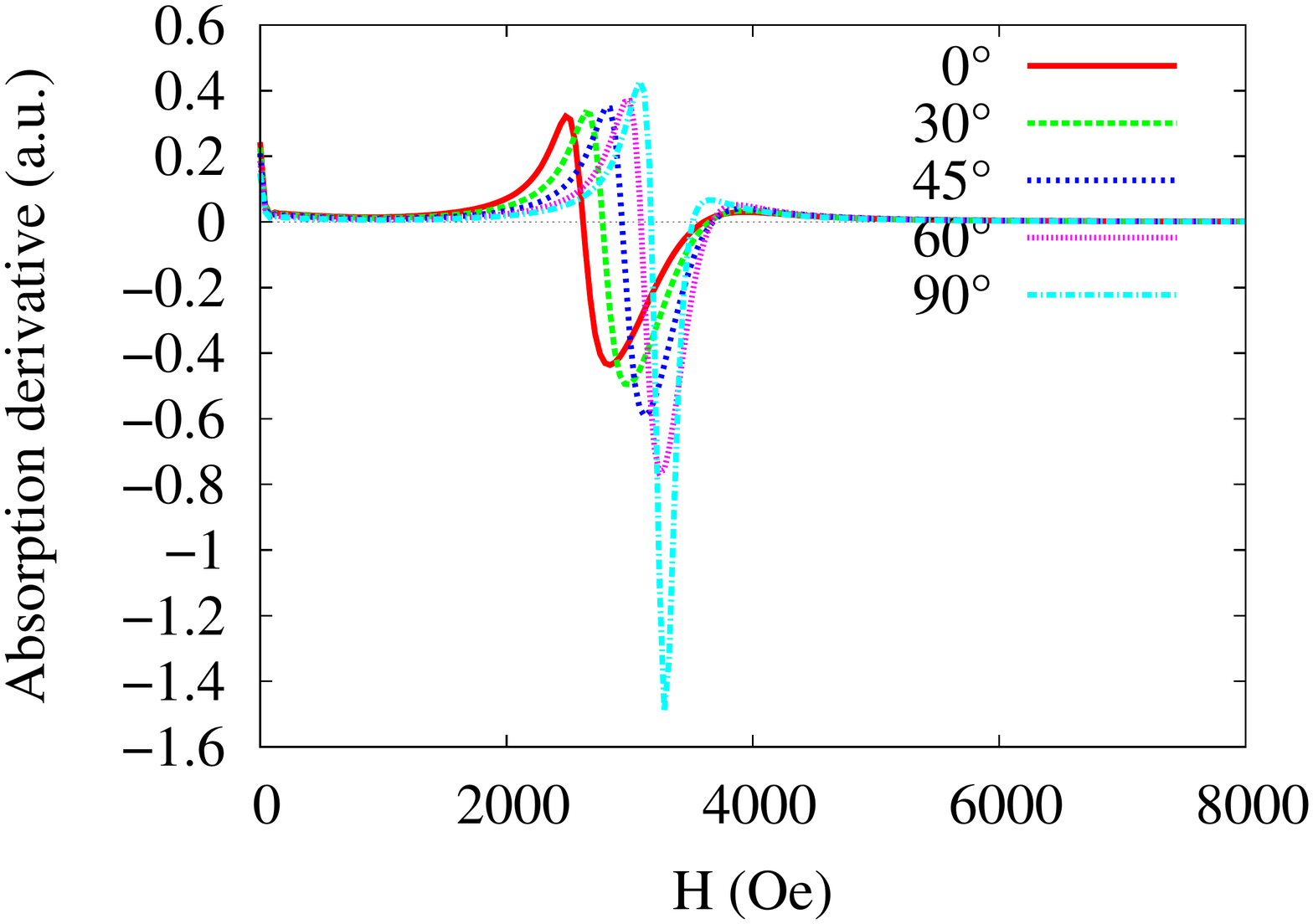}} \\ 
      \resizebox{80mm}{!}{\includegraphics[angle=0]{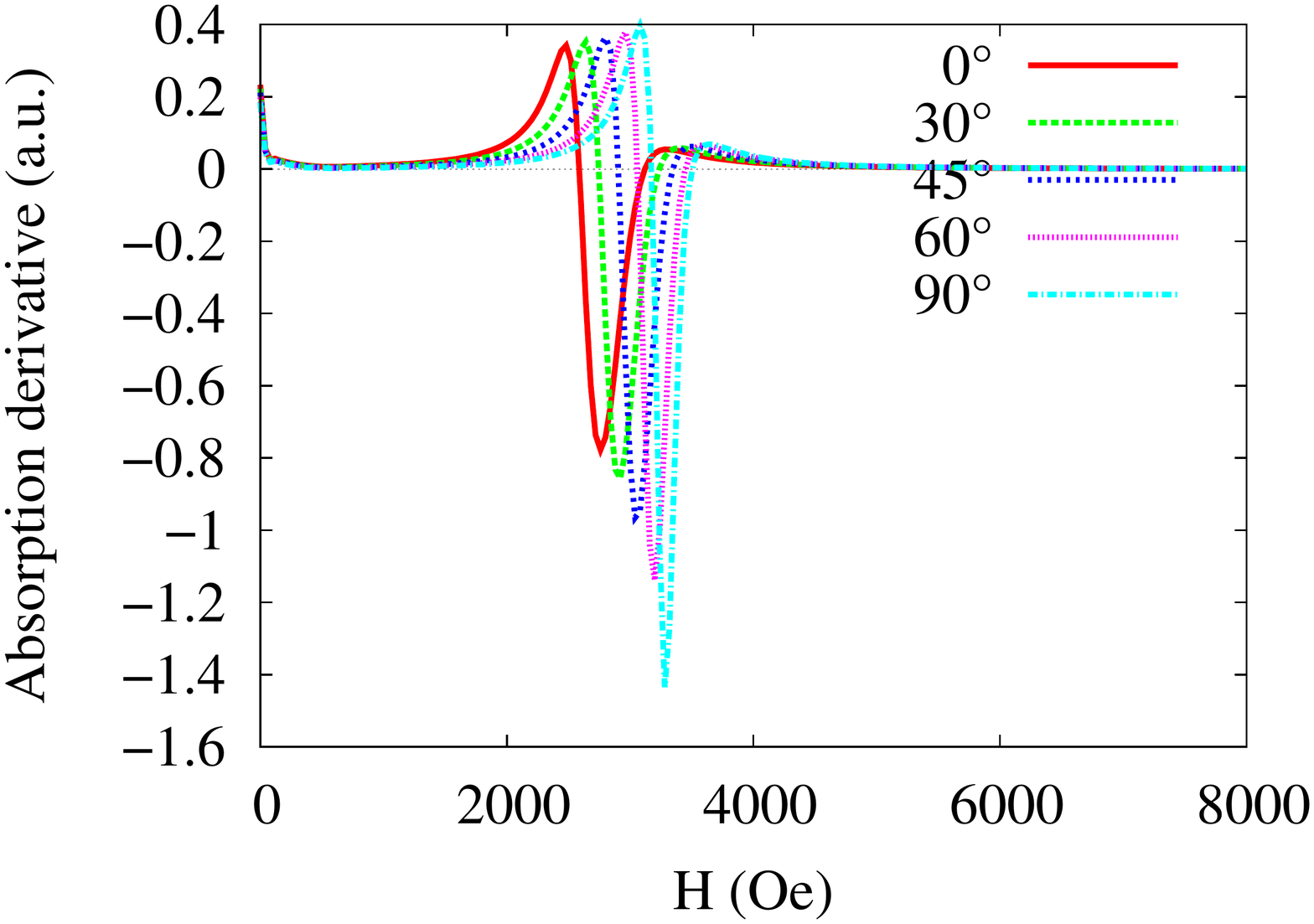}} \\
      \resizebox{80mm}{!}{\includegraphics[angle=0]{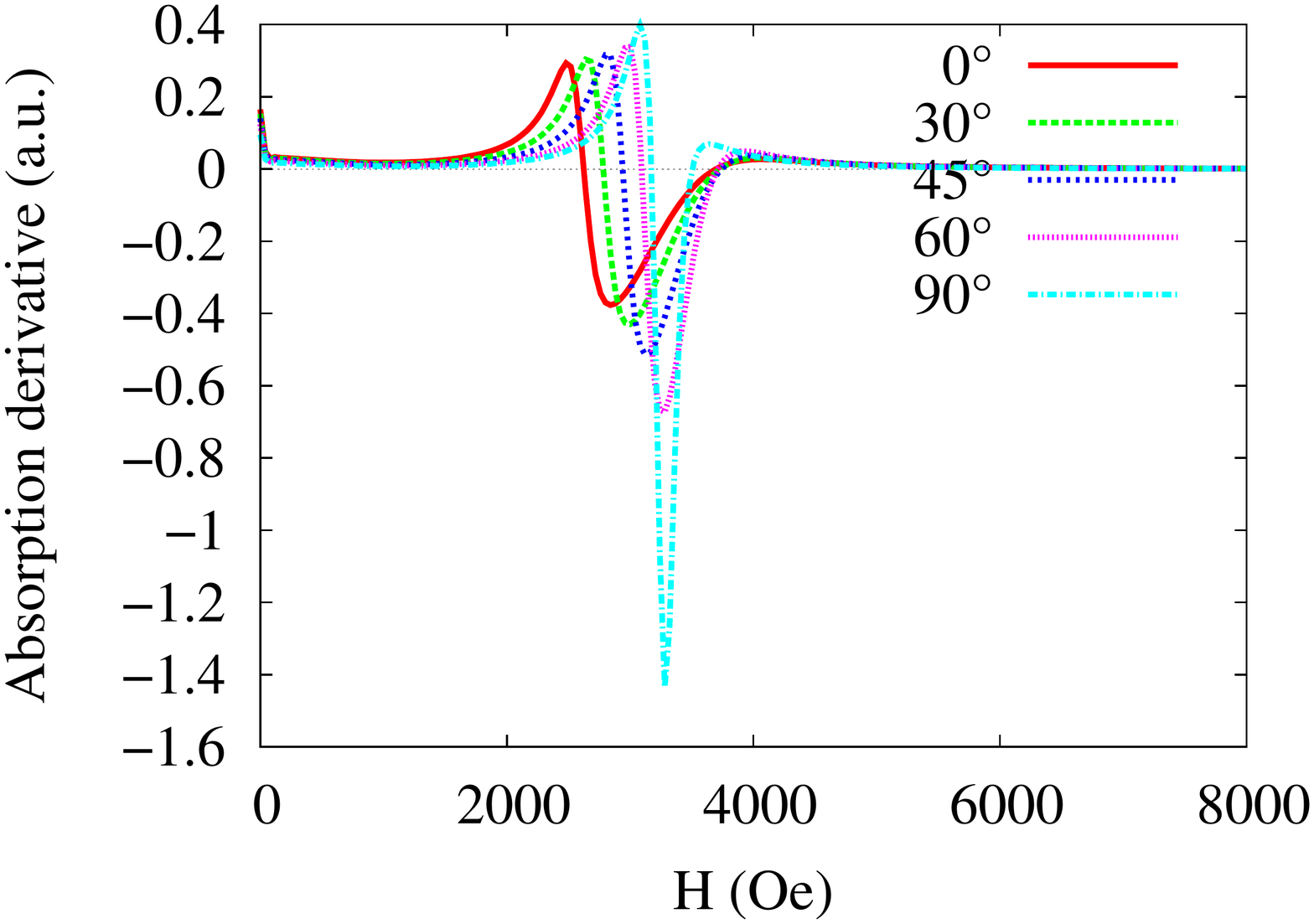}} \\  
    \end{tabular}
\caption{(Color on-line) Calculated derivative of the absorption 
$\frac{d <\chi_{xx}^{''}>}{dH}$ versus field for several field angles
$\theta_H$ and all diameters. The Preisach parameters are the same used in the hysteresis
loop fit. The lineshape is in arbitrary units and drawn for field angles of
0, 30, 45, 60 and  90 degrees, in all diameter cases: 15, 50, 80 and 100 nm (from top to bottom).}
\label{chidh}
\end{center}
\end{figure}

Performing the above double integral over the Preisach plane we spline the values obtained and
take the derivative with respect to $H$ from the splined value (see for instance Numerical Recipes~\cite{NR}).
The results are displayed in fig.~\ref{chidh}.

\begin{figure}
\begin{center}
    \begin{tabular}{c}
      \resizebox{80mm}{!}{\includegraphics[angle=0]{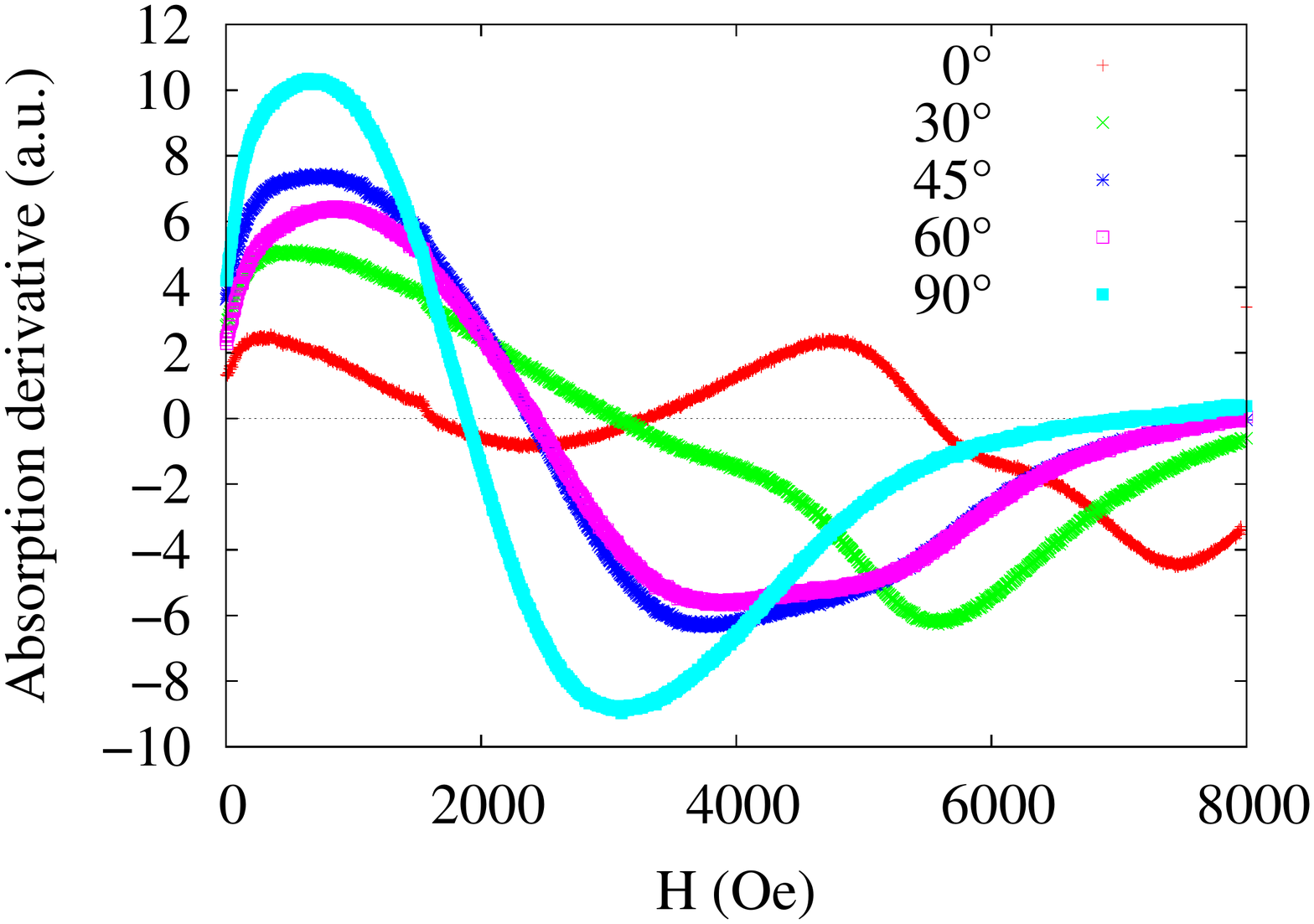}} \\
      \resizebox{80mm}{!}{\includegraphics[angle=0]{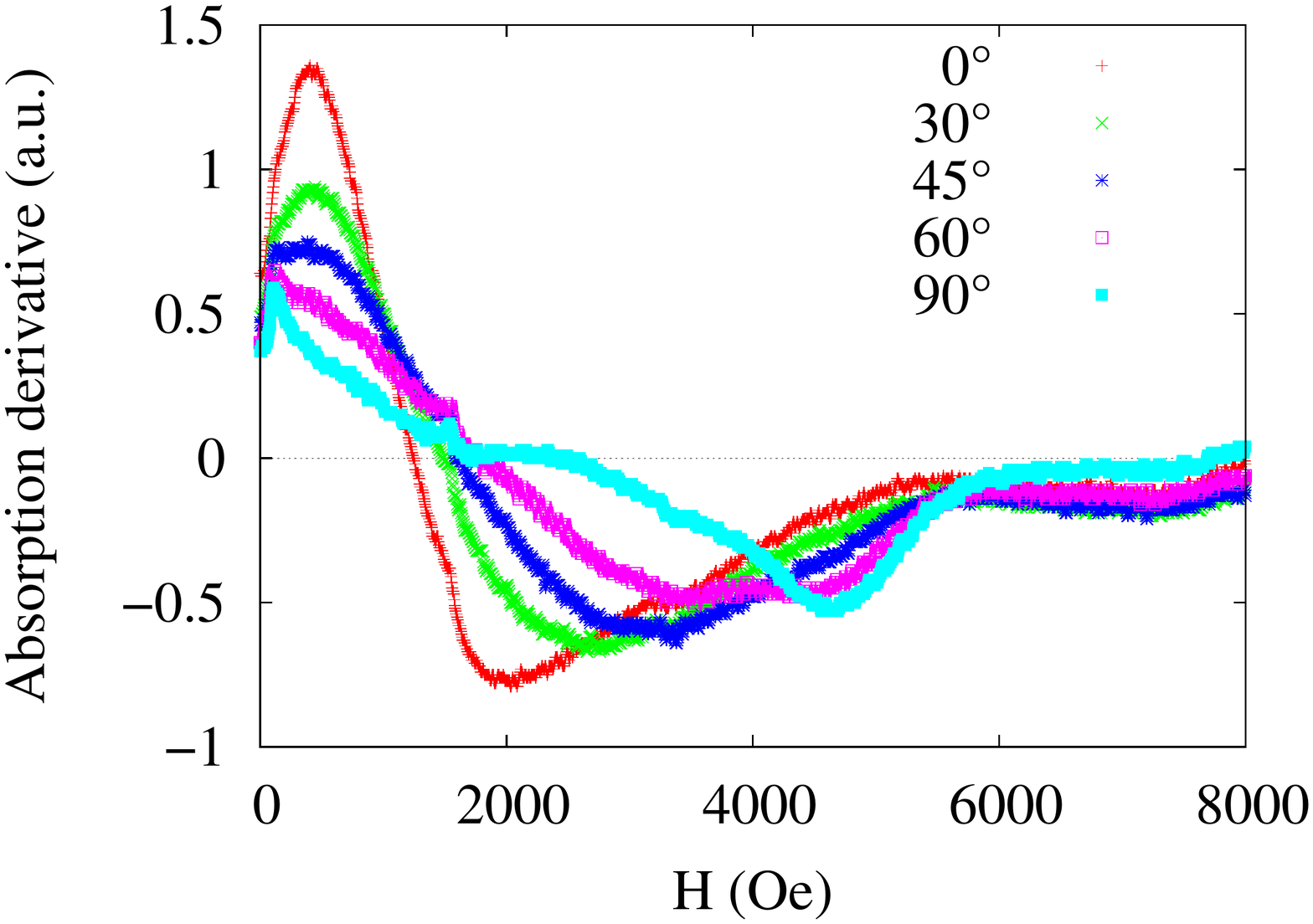}} \\ 
      \resizebox{80mm}{!}{\includegraphics[angle=0]{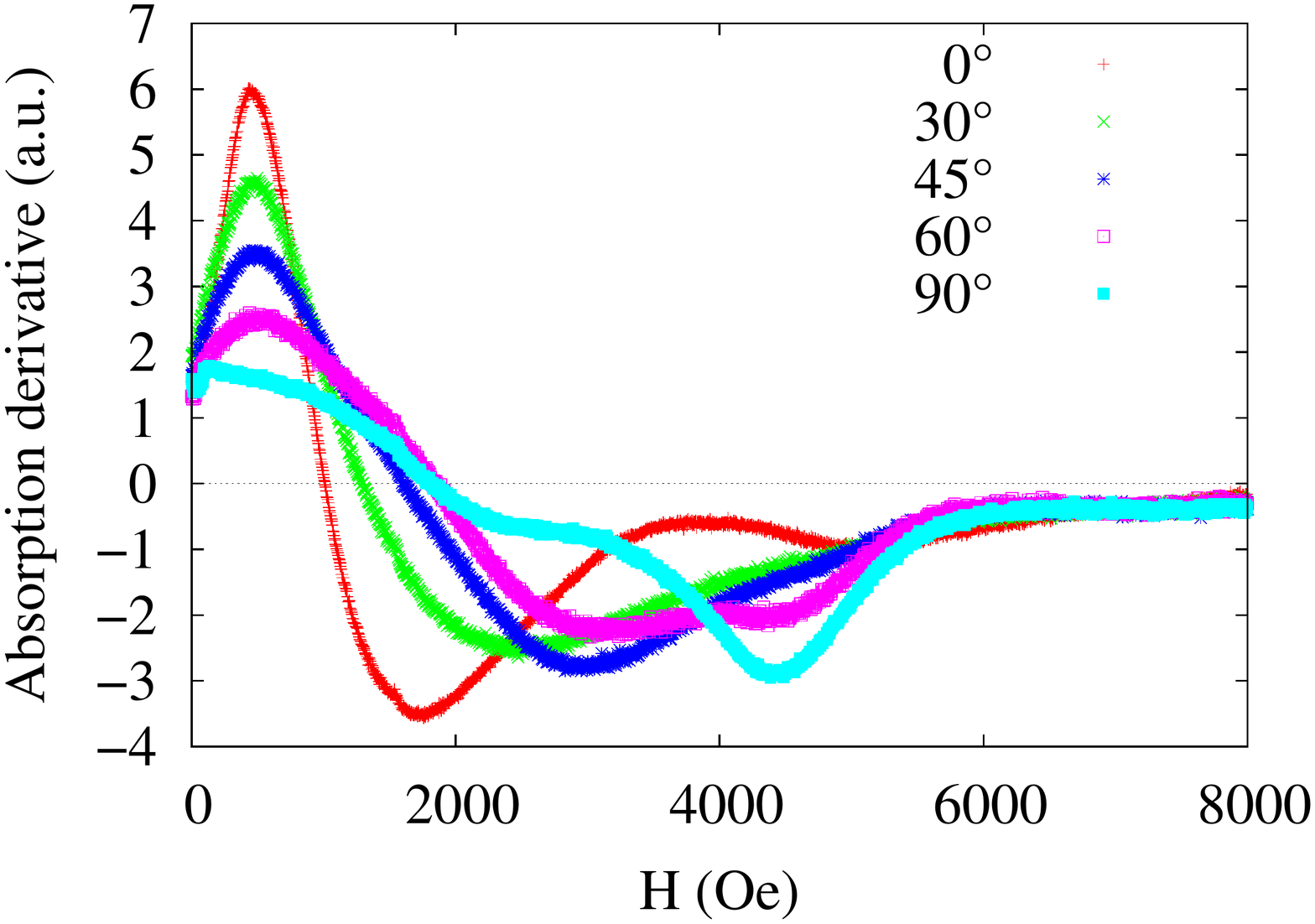}} \\
      \resizebox{80mm}{!}{\includegraphics[angle=0]{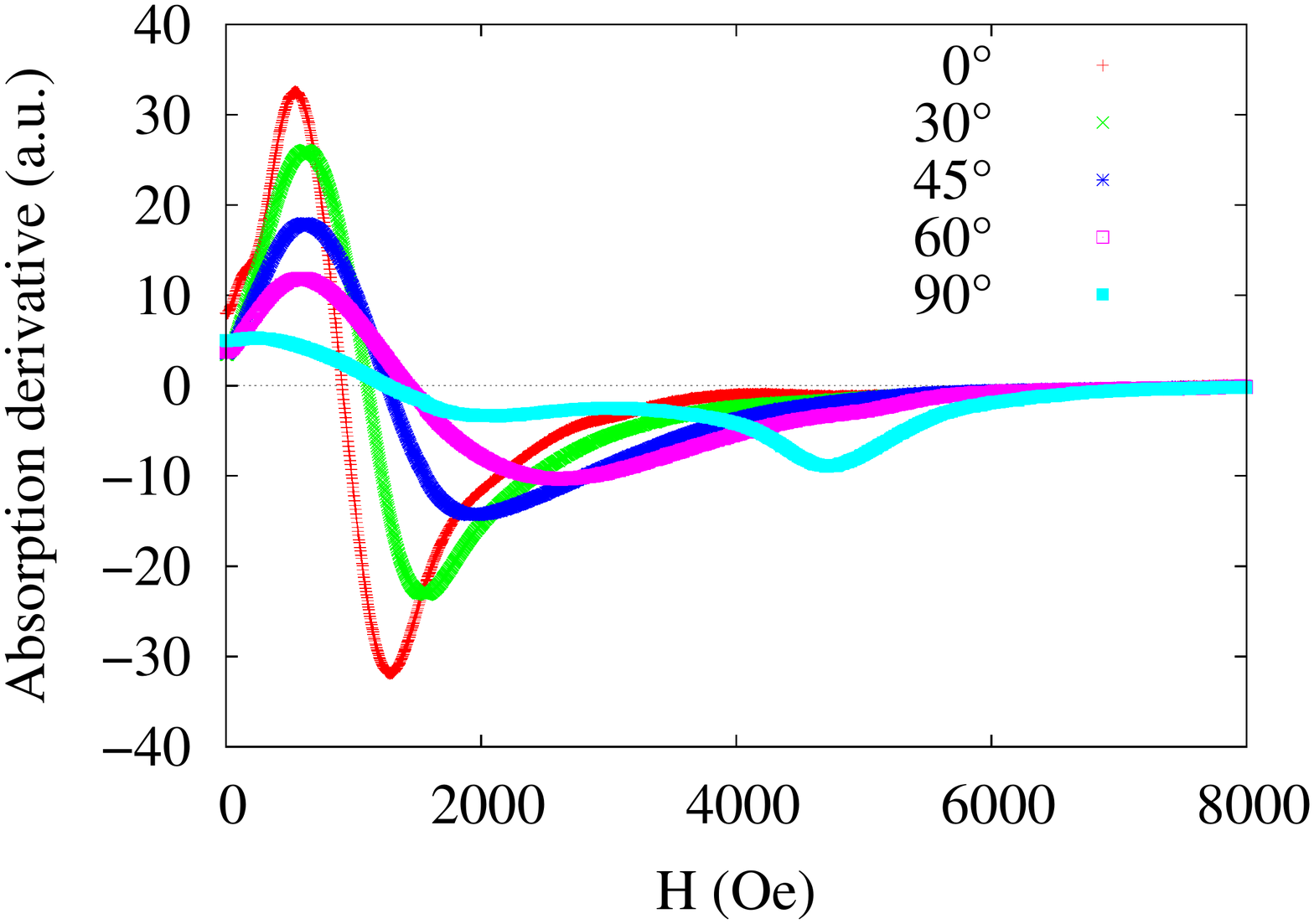}} \\  
    \end{tabular}
    \caption{(Color on-line) Measured FMR lineshape versus field for different angles with respect to nanowire axis  
at a frequency of 9.4 GHz and at room temperature. The lineshape $\frac{d <\chi_{xx}^{''}>}{dH}$ in
arbitrary units is drawn for various angles $\theta_H=$ 0, 30, 45, 60 and  90 degrees, 
for 15, 50, 80 and 100 nm nanowire diameters (from top to bottom).}
    \label{lineshapes}
\end{center}
\end{figure}

The FMR derivative spectrum is easily found to be asymmetric in contrast to what is normally
obtained with Landau-Lifshitz-Gilbert modeling. The Preisach PM2 results agree with lineshapes 
previously obtained in the literature by Ebels \etal~\cite{Ebels} as well as Dumitru \etal~\cite{Dumitru} 
but not with our measurements (see fig.~\ref{lineshapes}) that display a small field shift with angle $\theta_H$. 

In table~\ref{Dumitru} we display results we obtain for the PM2 parameters that fitted
the hysteresis loops and FMR measurements of $\chi_{xx}^{''}$ versus field in the 
Ni2 and Ni6 sample cases~\cite{Dumitru}. Note that some
values of Table~\ref{Dumitru} are different from those given in Table I of ref.~\cite{Dumitru}, 
nevertheless it shows that the PM2 model is capable of achieving hysteresis loop and
FMR results for the samples Ni2 and Ni6.

\begin{table}[htbp]
\begin{center}
\caption{Results obtained for PM2 model parameters belonging to samples Ni2 and Ni6 studied by 
Dumitru \etal~\cite{Dumitru}. We differ from some of the parameters displayed in their Table I.}
\begin{tabular}{|c|c|c|c|}
\hline
\hline
Fields (Oe) & Field orientation &  Ni2 & Ni6   \\
\hline
$H_0$ & in wire plane & 120 & 180  \\
 & perpendicular to wire plane & 125 & 170  \\
\hline
$h_{i0}$ & in wire plane & 1430 & 180  \\
 & perpendicular to wire plane & 70 & 580  \\
\hline
$\sigma_i$ & in wire plane & 260 & 720  \\
 & perpendicular to wire plane & 250 & 610  \\
\hline
$\sigma_c$  & in wire plane & 40 & 240  \\
 & perpendicular to wire plane & 60 & 265  \\
\hline
\hline
\end{tabular}
\label{Dumitru}
\end{center}
\end{table}

Turning to the calculated linewidths concerning our FMR measurements, we infer that they
are smaller than the experimental values which implies that we have
to include additional interactions in the dynamic (FMR) calculation. 
This is due to the fact we concentrate on the PM2 model with the same values of the parameters 
that previously fitted the hysteresis loops for all field angles. This contrasts with
Dumitru \etal~\cite{Dumitru} who did the fit for two orientations of the field only ($\theta_H=$ 0\deg and 90\deg ).

Moreover, let us point out from the $H_{res}$ versus $\theta_H$ fit (in Table~\ref{fit_table})
that in the dynamic case, several values, such as the anisotropy constant $K_1$ changes
significantly, when diameter is reduced from 100 nm to 15 nm because of the appearance of surface
anisotropy~\cite{Tannous}. Therefore a more complex Preisach model needed in order to fit
static and dynamic results with the same sets of parameters for all angles.  

\subsection{Angular Analysis of FMR lineshape width results}

Measured FMR lineshapes (absorption derivative spectra~\cite{Ebels})  
for different angles are displayed in fig.~\ref{lineshapes} for
15 nm, 50 nm, 80 nm and 100 nm diameter cases. Generally the lineshapes
behave versus magnetic field as the derivative of a Lorentzian. 

We use a least-squares algorithm to extract the values of the Lorentzian 
widths as explained in Appendix I.

The width results displayed in Table~\ref{Width} show that it is possible to
relate unambiguously the value of the Lorentz derivative width to the angle $\theta_H$ for
a given diameter, hence the possibility to build angle sensors on the basis of that
observation.

\begin{table}[htbp]
\begin{center}
\caption{Lorentz derivative widths (in Oe), versus field angle $\theta_H$, fitted with respect to
experimental angular FMR lineshapes pertaining to 15, 50, 80 and  100 nm diameter samples.}
\begin{tabular}{|c|c|c|c|c|c|}
\hline
$d$(nm) & 0\deg &  30\deg & 45\deg & 60\deg & 90\deg\\
\hline
15 & 3557 & 2742 & 1806 & 1791 & 1364 \\ 
50 & 1020 & 1120 & 1251 & 1495 & 2069  \\
80 & 839 & 1092 & 1278 & 1349 & 2215  \\ 
100 & 638 & 777 & 909 & 1100 & 1227 \\ 
\hline
\end{tabular} 
\label{Width}
\end{center}
\end{table}

Lorentz derivative widths generally decrease as we increase nanowire diameter for all angles.
For a fixed diameter, they increase with angle in the 50, 80 and 100 nm whereas in the
15 nm case they decrease. This originates from the fact, the 15 nm case possesses a large 
surface anisotropy as analysed previously in ref.~\cite{Tannous}, besides the lineshape
behavior in this case is more complicated than the larger diameter cases.

\section{Discussion and Conclusion}
We have performed angular Preisach analysis for the static (VSM hysteresis loops)
and dynamic measurements (FMR lineshape widths) and shown that many results 
can be deeply understood and might be further developed in order to be embedded 
in applications such as angle detection sensors.

The transition at 50 nm in VSM and FMR measurements is extremely promising 
because of several potential applications in race-track MRAM devices. 
Yan \etal~\cite{Yan} predicted that in Permalloy nanowires of 50 nm and less, 
moving  zero-mass domain walls may attain a velocity of several 100 m/s 
beating Walker limit obeyed in Permalloy strips with same lateral size.  
Hence, nanowire cylindrical geometry in contrast to prismatic geometry of stripes
bears important consequences on current injection in nanowires
that applies Slonczewski type torques~\cite{Kiselev} on magnetization affecting 
domain wall motion with reduced Ohmic losses~\cite{Tretiakov}.

Ordered arrays of nanowires are good candidates for patterned
media and may also be used in plasmonic applications 
such as nano-antenna arrays or nanophotonic waveguides in integrated 
optics~\cite{plasmon}. Recently~\cite{Stipe}, heat assisted magnetic perpendicular
recording using plasmonic aperture nano-antenna has been tested on patterned
media in order to process large storage densities starting at 1 Tbits/in$^2$
and scalable up to 100 Tbits/in$^2$.

While the Preisach PM2 model can explain separately the static or dynamic 
results, one might extend it through the use of other distributions
of interaction and coercivity in order to explain  the VSM and FMR
measurements simultaneously for all angles. 

Nonetheless, the angular behavior of the linewidth is
interesting enough to consider its use in angle sensors that might compete with 
present technology based on AMR or GMR effects.

\section*{Acknowledgments}
Some of the FMR measurements were kindly made by Dr. 
R. Zuberek at the Institute of Physics of the Polish Academy of Science, Warsaw (Poland).

\appendix

\section{Angular FMR linewidth evaluation procedure}

We have developed a procedure based on a least squares minimization procedure of the curve
$\frac{d\chi^{''}}{dH}$ versus $H$ to the set of $n$ experimental measurements ${[x_{i},~y_{i}]}_{i=1,n}$ where
$x_i=H_{i}$ and $y_{i}= \frac{d\chi^{''}(\bm{ \beta}; x_{i})}{dH}$. $\bm{ \beta}$ represents a set 
of fitting parameters. 

One of the parameters is the width $\Delta H$ obtained from a fitting procedure to the derivative of a Lorentzian whose expression is given by:
\begin{equation}
\left[\frac{d\chi^{''}}{dH}\right]_{\mathscr{L}}=\frac{A (H-H_0)}{({\Delta H}^2 + (H-H_0)^2)^2}
\end{equation}

$H_0$ is not a fitting parameter since it can be determined by the intersection of the lineshape
with the $H$ axis. The parameters  $A, \Delta H$ are determined with a fitting procedure.
Writing the set of minima equations to be satisfied at the data points:
\begin{equation}
\frac {1}{n}\sum_{i=1}^{n}\left\{\left[\frac{d\chi^{''}(\bm{ \beta}; x_{i})}{dH}\right]_{\mathscr{L}}- y_{i} \right\}^{2} \mbox{  minimum},  \nonumber \\
\label{eqfit}
\end{equation}

The fitting method is based on the Broyden algorithm, a generalization to
higher dimension of the one-dimensional secant method \cite{NR} that allows us to determine 
in a least-squares fashion, the set  of unknowns $A, \Delta H$. Broyden method
is selected because it can handle over or under-determined numerical problems and that
it works from a singular value decomposition point of view \cite {NR}.  This means it 
is able to circumvent singularities and deliver a practical solution to the problem at hand
as an optimal set\cite{NR} within a minimal distance from the real one. 

\section{Preisach formalism overview}

Preisach model~\cite{mayergoyz91} is based on a statistical approach towards
magnetization processes \cite{mayergoyz91}, \cite{bertotti98}.

Comparing Preisach formalism and micromagnetics is akin to understanding the link
between thermodynamics and statistical mechanics.

The nanowire array is viewed as made of single domain  interacting entities 
each nanowire being represented by a switching field and a local 
interaction field.  The local interaction field in each nanowire is 
assumed to be constant. A system of interacting nanowires is represented by a 
probability density function (PDF) $p(h_i,h_{c})$ depending on interaction 
and coercive fields $H_{i}$ and $H_{c}$ respectively.

The main objective in Preisach modeling is to find the best  $p(h_{i},h_{c})$ such that
the best possible agreement with system behavior is obtained. 

Preisach formalism is an energy-based description of hysteresis, and does not 
require that the material under investigation be decomposable into discrete physical
entities such as magnetic particles.

It assumes that the magnetic system free energy functional can be decomposed into an ensemble of 
elementary two level (double well) subsystems (TLS)\cite{mayergoyz91}. 

\begin{figure}
  \begin{center}
    \begin{tabular}{c}
      \resizebox{60mm}{!}{\includegraphics{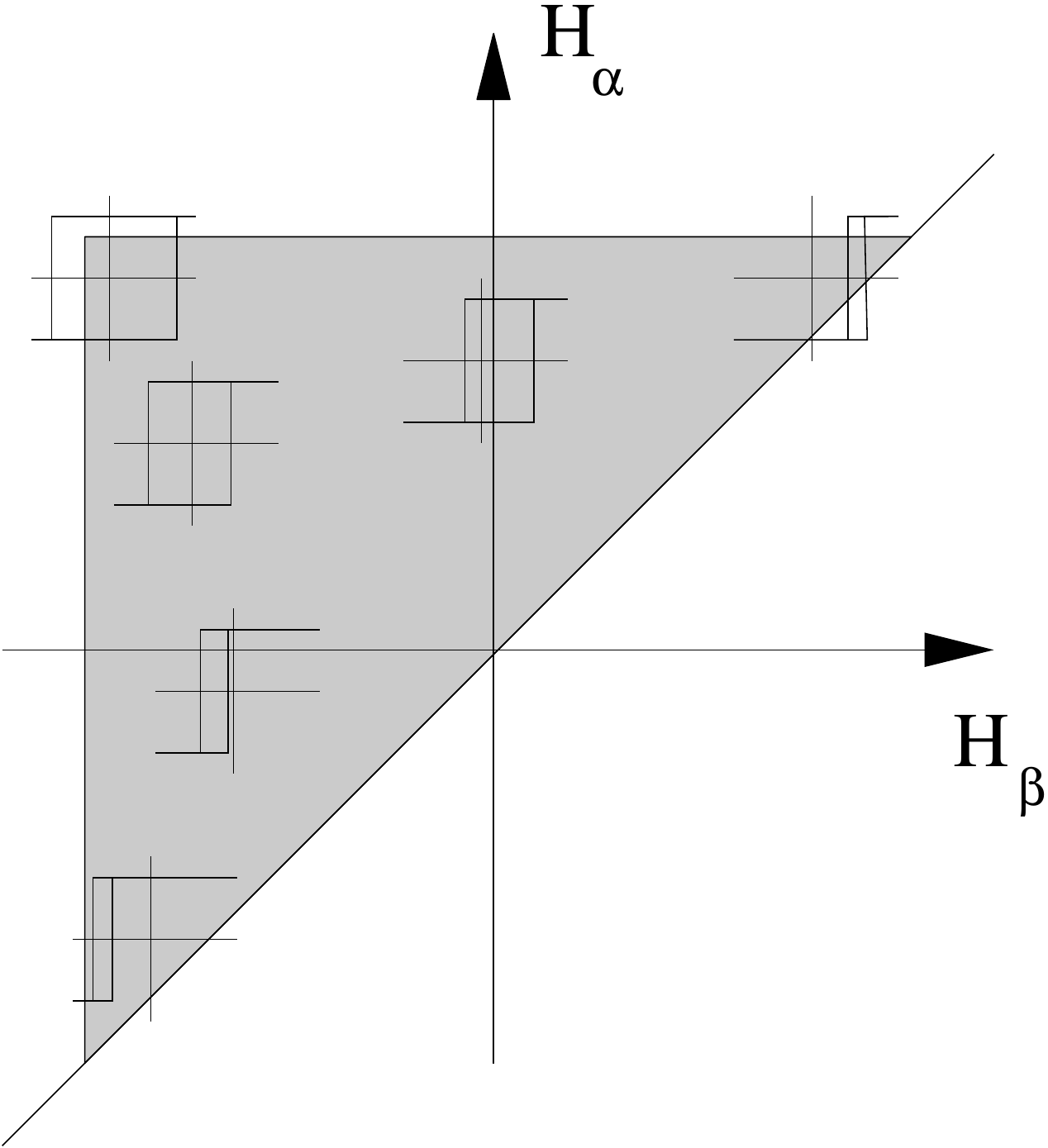}} \\
      \resizebox{60mm}{!}{\includegraphics{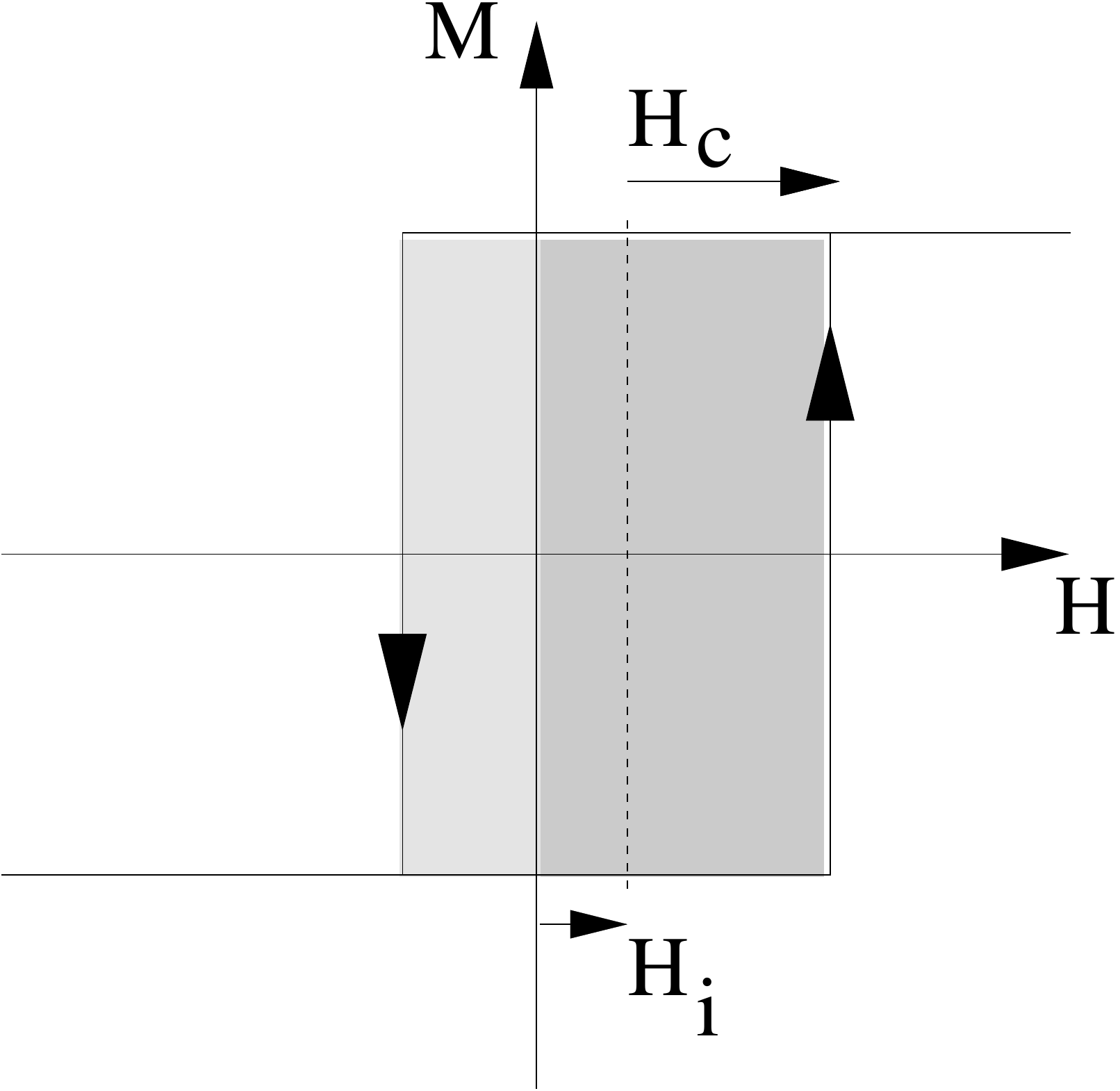}} 
    \end{tabular}
    \caption{(a) Preisach ($H_\alpha, H_\beta$) plane displaying different hysteresis loops 
at different locations. At each point an elementary hysteresis loop is displayed 
characterized by the values of the interaction $H_i$ and coercive $H_c$ fields. 
(b) An elementary hysteron with a square hysteresis loop 
is shown shifted by the interaction field $H_i$ and possessing a half-width given by the 
coercive field $H_c$.}
    \label{pplane}
  \end{center}
\end{figure}

It represents the magnetic material into a collection of microscopic bistable units
"hysterons" having statistically distributed 
coercive and interaction fields. Each unit is characterized by a rectangular hysteresis 
loop (see fig.~\ref{pplane}) and its status is determined by the actual field and history
of the applied external fields.  

The classical version (CPM) of the model is based on the use of a joint distribution of
normalized interaction fields $H_i$ and coercive fields $H_c$. The interaction fields $H_i$ induce a 
shift in the elementary hysteresis loop (see fig.~\ref{pplane}) whereas the coercive fields
increase its width. Integrating the density over a given
path in the Preisach plane yields a magnetization process. 

These fields originate from the existence of switching fields $(H_\alpha, H_\beta)$ that span the Preisach 
plane (see fig.~\ref{pplane}) such that:

\begin{eqnarray}
H_\alpha =  H_i + H_c, \hspace{0.2cm} H_\beta = H_i - H_c,  \nonumber \\
 H_i = \frac{H_\alpha + H_\beta}{2} , \hspace{0.2cm} H_c = \frac{H_\alpha - H_\beta}{2}
\label{hplane}
\end{eqnarray}

Mayergoyz~\cite{mayergoyz91} has demonstrated that the
necessary and sufficient conditions for a system to be rigorously described by a CPM 
are the wiping-out and congruency properties~\cite{Stancu}. A hysteretic system will present 
the wiping-out property when it returns to the same state after performing a minor loop. 
The second property refers to the shape of the minor loops measured in the same field range; 
if all these minor loops are congruent within a given field range and this property 
does not depend on the actual field range used in the experiment, the system obeys the congruency property.

The major hysteresis loop
is obtained when a path linear in the applied field is used \cite{mayergoyz91}, \cite{bertotti98}.
Thus, in order to estimate the the hysteresis loop, we determine the magnetization $M$ 
with a double integration over the PDF as given below:

\begin{equation}
M=2 M_S \int_0^\infty dh_i \int_0^{b(h_i)} dh_c p(h_i, h_c) 
\label{magnetization}
\end{equation}

where the function $b(h_i)$ represents such path. 

The $H_i,H_c$ PDF can be either analytically
built from standard PDF (Gaussian, Lorentzian, uniform etc...) or experimentally
determined from FORC (First Order Reversal Curves) measurements.
This originates from  the fact magnetic interactions between nanowires are a 
major determinant of noise levels  in magnetic media whether it is used for storage or 
processing such as in spintronics. 

While conventional methods of characterizing magnetic
interactions utilize Isothermal Remanent Magnetization (IRM) and remanence DC Demagnetization (DCD)
curves~\cite{mayergoyz91}, Preisach modeling is based on some distribution 
which is supposed to adequately describe the magnetic system at hand.

FORC is a popular measurement leading to an appropriate Preisach model. 
It begins with sample saturation  with a large positive field. 
The field is ramped down to a reversal field $H_\alpha$. FORC consists of 
a measurement of the magnetization as the field is then increased from $H_\alpha$  back 
up to saturation. The magnetization at applied field $H_\beta$ on the FORC with 
reversal point $H_\alpha$ is denoted by $M (H_\alpha, H_\beta )$, where $ H_\beta  \geq H_\alpha$ . 

The PDF is obtained from the second mixed derivative:

\begin{equation}
p(H_\alpha, H_\beta)=-\frac{\partial^2 M}{ \partial H_\alpha \partial H_\beta}
\label{FORC}
\end{equation} 

Let us assume we adopt a double Lorentzian PDF given by:
\begin{eqnarray}
p(h_i,h_c)=\frac{2}{ \pi \sigma^2_i H^2_0 [\frac{\pi}{2}+\tan^{-1}(\frac{1}{\sigma_i}) ] } \times \nonumber \\
\frac{1}{ [ 1+(\frac{h_i+h_c-H_0}{\sigma_i H_0})^2 ] [ 1+(\frac{h_i-h_c-H_0}{\sigma_i H_0})^2 ]}
\label{LORENTZ}
\end{eqnarray}

with $h_i,h_c$ a set of normalized fields,  $\sigma_i, \sigma_c$
the standard deviation of the individual Lorentzian/Gaussian distributions considered as independent)
we find hysteresis loops that are upstraight
whereas the VSM measured loops exhibit some inclination.

%\begin{figure}[htbp]
%\centerline{\includegraphics[angle=0,width=3.5in]{loopCPM1.pdf}}
%\caption{Example Preisach hysteresis loop for the double Lorentzian model with the parameters 
%$\sigma_i=0.1, H_0=1$.}
%\label{CPMloop1}
%\end{figure}

When one uses rather a double Gaussian PDF as in the CPM case, we get inclined hysteresis loops
as observed with the VSM measurements.

%\begin{figure}[htbp]
%\centerline{\includegraphics[angle=0,width=3.5in]{loopCPM2.pdf}}
%\caption{Example Preisach hysteresis loop for the double Gaussian model with the parameters 
%$\sigma_i=10., \sigma_c=10., H_0=1$.}
%\label{CPMloop2}
%\end{figure}

Nevertheless both approaches do not agree with the hysteresis loops that we find
experimentally as described by Dumitru \etal~\cite{Dumitru}. This is why we use the PM2 model
as explained in section \ref{Preisach}.


\begin{thebibliography}{11}
\bibitem{coupling} Coupling  GHz range  electromagnetic waves (with $\lambda \sim $ 30 cm) to nanowires of diameter $d \sim $100 nm must respect the cutoff condition $\lambda \le d$. 
Devices using surface plasmons  
may be used to inject electromagnetic energy through nanowires (see for instance Ditlbacher \etal~\cite{plasmon}).
\bibitem{cubic} In spite of the fact bulk Nickel has cubic anisotropy~\cite{aniso},
our description based on uniaxial anisotropy is sufficient~\cite{Wegrowe} to describe 
the FMR spectra. For a description based on a cubic anisotropy see De La Torre Medina \etal~\cite{Darques}.
\bibitem{aniso} Cubic anisotropy energy to sixth-order is:
$K_1 (\alpha_1^2 \alpha_2^2 +\alpha_2^2 \alpha_3^2 +\alpha_3^2 
\alpha_1^2 )+K_2 \alpha_1^2 \alpha_2^2 \alpha_3^2$
where: $\alpha_1, \alpha_2, \alpha_3$ are the cosines of the angles, the magnetization
makes with the $(\bm{x},\bm{y},\bm{z})$ axes respectively (see fig.~\ref{coord}).
At room temperature, bulk Nickel values are: 
$K_1=-4.5 \times 10^{4}$ erg/cm$^3$, $K_2=2.3 \times 10^{4}$ erg/cm$^3$ and $M_S$=485 emu/cm$^3$.
\end{thebibliography}
\end{document}